# Tracking Social Media Discourse About the COVID-19 Pandemic: Development of a Public Coronavirus Twitter Data Set

Emily Chen, Kristina Lerman, Emilio Ferrara
*University of Southern California, Information Sciences Institute*

## Abstract

**Background:**
At the time of this writing, the novel coronavirus (COVID-19) pandemic outbreak has already put tremendous strain on many countries' citizens, resources and economies around the world. Social distancing measures, travel bans, self-quarantines, and business closures are changing the very fabric of societies worldwide. With people forced out of public spaces, much conversation about these phenomena now occurs online, e.g., on social media platforms like Twitter.

**Objective:**
In this paper, we describe a multilingual coronavirus (COVID-19) Twitter dataset that we are making available to the research community via our COVID-19-TweetIDs Github repository.

**Methods:**
We started this ongoing data collection on January 28, 2020, leveraging Twitter's Streaming API and Tweepy to follow certain keywords and accounts that were trending at the time the collection began, and used Twitter's Search API to query for past tweets, resulting in the earliest tweets in our collection dating back to January 21, 2020.

**Results:**
Since the inception of our collection, we have actively maintained and updated our Github repository on a weekly basis. We have published over 123 million tweets, with over 60% of the tweets in English. This manuscript also presents basic analysis that shows that Twitter activity responds and reacts to coronavirus-related events.

**Conclusions:**
It is our hope that our contribution will enable the study of online conversation dynamics in the context of a planetary-scale epidemic outbreak of unprecedented proportions and implications. This dataset could also help track scientific coronavirus misinformation and unverified rumors or enable the understanding of fear and panic – and undoubtedly more.

## Introduction

The first cases of a novel coronavirus disease (officially named COVID-19 by the World Health Organization (WHO) on February 11, 2020) were reported in Wuhan, China in late December 2019; the first fatalities were reported in early 2020. The fast-rising infections and death toll led the Chinese government to quarantine the city of Wuhan on January 23, 2020 [1]. During this period, other countries began reporting their first confirmed cases of the disease, and on January 30, 2020 WHO announced a Public Health Emergency of International Concern. With more countries reporting cases of the disease, and infections rapidly escalating in some regions of the world, including South Korea, Iran and Italy, the WHO has upgraded COVID-19 to a pandemic [2].

As of the time of this writing, COVID-19 has been reported in 185 countries, leaving governments all over the world scrambling for ways to contain the disease and lessen its adverse consequences to their people's health and economy [3].

Preventative measures implemented by national, state and local governments now affect the daily routines of millions of people worldwide [4]. "Social distancing," the most widely used of such measures, aims to curtail new infections by reducing physical contact between people [5]. Social distancing measures have led to the cancellation of sporting events and conferences [6], closures of schools and colleges [7], and has forced many businesses to require their employees to work from home [8]. As more and more social interactions move online, the conversation around COVID-19 has continued to expand, with growing numbers turning to social media for both information and company [9, 10]. Platforms such as Twitter have become central to the technological and social infrastructure that allows us to stay connected even during crises.

We describe a Twitter dataset about COVID-19-related online conversations that we are sharing with the research community. People all over the world take to Twitter to express opinions and engage in dialogue in a public forum, and, with Twitter's open API, has proved to be an invaluable resource in studying a wide range of topics. Twitter has long been used by the research community as a means to understand dynamics observable in online social networks: from information dissemination [11, 12], to the prevalence and influence of bots and misinformation [13,14]. More importantly during the current coronavirus pandemic, Twitter provides researchers the ability to study the role social media plays in the global health crisis [15-19]. We hope that this data will spur new research about the social dimensions of the pandemic.

We began collecting data in real-time from Twitter, with the earliest tweets from January 21, 2020, tracking COVID-19 related keywords and accounts. Here we describe the data collection methods, document initial data statistics, and provide information about how to obtain and use the data.

## Methods

We began actively collecting tweets from January 28, 2020, leveraging Twitter's streaming API [20] and Tweepy [21] to follow specific keywords and accounts that were trending at the time. When we started collecting tweets, we also used Twitter's search API [22] on the same keywords to gather related historical tweets. Thus, the earliest tweets in our collection date back to January 21, 2020. Since then, we have incrementally added keywords and accounts to follow based on the conversations occurring on Twitter at any time. We have collected over 72 million tweets from the inception until March 21, 2020, constituting roughly 600 GB of raw data, and are still collecting data to this day.

Our collection relies upon publicly available data and is hence registered as IRB exempt by USC IRB (approved protocol UP-17-00610). We release the dataset with the stipulation that those who use it must comply with Twitter's Terms and Conditions [23].

## Tracked Keywords and Accounts

By continuously monitoring Twitter's trending topics, keywords and sources associated with COVID-19, we did our best to capture conversations related to the coronavirus outbreak.

Twitter's streaming API returns any tweet containing the keyword(s) in the text of the tweet, as well as in its metadata; therefore, it is not always necessary to have each permutation of a specific keyword in the tracking list. For example, the keyword "Covid" will return tweets that contain both "Covid19" and "Covid-19". We list a subset of the keywords and accounts that we are following in Table 1 and Table 2 respectively, along with the date we began tracking them. There are some keywords that overlap due an included keyword being a sub-string of another, but we included both for good measure. The keyword choices in the current dataset are all in English, so there is a heavy bias towards English tweets and events related to English speaking countries. Due to the evolving nature of the pandemic and online conversations, these tables will expand as we continue to monitor Twitter for additional keywords and accounts to add to our tracking list.

**Table 1**. A sample of the keywords that we are actively tracking in our Twitter collection. Please see the Github repository for a full list of all tracked keywords (v1.8 — May 8, 2020)

| Keyword | Tracked Since | Keyword | Tracked Since |
|---|---|---|---|
| Coronavirus | 1/21/2020 | CancelEverything | 3/13/2020 |
| Corona | 1/21/2020 | Coronials | 3/13/2020 |
| CDC | 1/21/2020 | SocialDistancing | 3/13/2020 |
| Ncov | 1/21/2020 | Panic buying | 3/14/2020 |
| Wuhan | 1/21/2020 | DuringMy14DayQuarantine | 3/14/2020 |
| Outbreak | 1/21/2020 | Panic shopping | 3/14/2020 |
| China | 1/21/2020 | InMyQuarantineSurvivalKit | 3/14/2020 |
| Koronavirus | 1/22/2020 | chinese virus | 3/16/2020 |
| Wuhancoronavirus | 1/22/2020 | stayhomechallenge | 3/16/2020 |
| Wuhanlockdown | 1/22/2020 | DontBeASpreader | 3/16/2020 |
| N95 | 1/22/2020 | lockdown | 3/16/2020 |
| Kungflu | 1/22/2020 | shelteringinplace | 3/18/2020 |
| Epidemic | 1/22/2020 | staysafestayhome | 3/18/2020 |
| Sinophobia | 1/22/2020 | trumppandemic | 3/18/2020 |
| Covid-19 | 2/16/2020 | flatten the curve | 3/18/2020 |
| Corona virus | 3/2/2020 | PPEshortage | 3/19/2020 |
| Covid | 3/6/2020 | saferathome | 3/19/2020 |
| Covid19 | 3/6/2020 | stayathome | 3/19/2020 |
| Sars-cov-2 | 3/6/2020 | GetMePPE | 3/21/2020 |
| COVID–19 | 3/8/2020 | covidiot | 3/26/2020 |
| COVD | 3/12/2020 | epitwitter | 3/28/2020 |
| Pandemic | 3/12/2020 | Pandemie | 3/31/2020 |
| Coronapocalypse | 3/13/2020 | | |

**Table 2**. Account names that we are actively tracking in our Twitter collection (v1.8 – May 8, 2020).

| Account Name | Tracked Since |
|---|---|
| PneumoniaWuhan | 1/22/2020 |
| CoronaVirusInfo | 1/22/2020 |
| V2019N | 1/22/2020 |
| CDCemergency | 1/22/2020 |
| CDCgov | 1/22/2020 |
| WHO | 1/22/2020 |
| HHSGov | 1/22/2020 |
| NIAIDNews | 1/22/2020 |
| DrTedros | 3/15/2020 |

## Results

### Releases

Our data collection will continue uninterruptedly for the foreseeable future. As the pandemic continues to run its course, we anticipate that the amount of data will grow significantly. The dataset is available on Github [24] and is released in compliance with the Twitter's Terms and Conditions, under which we are unable to publicly release the text of the collected tweets. We are, therefore, releasing the Tweet IDs, which are unique identifiers tied to specific tweets. The Tweet IDs can be used by researchers to query Twitter's API and obtain the complete Tweet object, including tweet content (text, URLs, hashtags, etc.) and authors' metadata. This process to retrieve the full Tweet object from Twitter starting from a Tweet ID is referred to as *hydration*. There are several easy to use tools that have been developed for such purposes, including the *Hydrator* [25] and *Twarc* [26], but one could also directly use Twitter's API to retrieve the desired data. This dataset can also be found on Harvard Dataverse [27]. Please see a list of basic statistics, including collection period and number of Tweets in that respective release, for all current (as of May 15, 2020) releases in Table 4.

There are a few known gaps in the data, which are listed in Table 5. Due to Twitter API restrictions on free data access, we were unable to recover the data from the listed times, as Twitter only provides free access to Tweets returned from their streaming API from the past week. To request access, interested researchers will need to agree upon the terms of usage dictated by the chosen license.

All of the Tweet ID files are stored in folders that indicate the year and month the Tweet was posted (YEAR-MONTH). The individual Tweet ID files each contain a collection of Tweet IDs, with the file names all beginning with the prefix "coronavirus-tweet-id-" followed by the year, month, date and hour the Tweet was posted (YEAR-MONTH-DATE-HOUR).

We note that if a Tweet has been removed from the platform, unfortunately researchers will not be able to obtain the original Tweet.

### *Most Recent Release (Release v1.8 – May 11, 2020)*

Our 9th release spans January 21, 2020 through May 8, 2020. The total dataset available now contains tweets from 22:00 UTC January 21, 2020 through 21:00 UTC May 8, 2020, with *123,113,914* tweets. The language breakdown of the tweets can be found in Table 4. A subset of the keywords and accounts that were followed during this time-frame can be identified by referencing Tables 1 and 2. For a full and up-to-date list of the keywords we are tracking, please see the "keywords.txt" file in the Github repository (a list of the accounts we are tracking can be found in the "accounts.txt" file). Some of the keywords may appear earlier than the initial listed track date in Table 1, as we systematically ran the same keywords through Twitter's search API to collect past instances of the keywords shortly after adding the keywords to be tracked in real-time.

*General Release Notes*

In order to use any Twitter facing libraries, including hydration software, users must first apply for a Twitter developer account and obtain the necessary authentication tokens [28].

The Github community has also generously contributed scripts to enable researchers to hydrate the Tweet IDs using *Twarc* [26].

**Table 3**. List of all releases and their statistics.

| Release Version | Release Date | Data Collection Period | Total # Tweets |
|---|---|---|---|
| v1.0 | 3/17/2020 | 3/05/2020 – 3/12/2020 | 8,919,411 |
| v1.1 | 3/23/2020 | 1/21/2020 – 3/12/2020 | 63,616,072 |
| v1.2 | 3/31/2020 | 1/21/2020 – 3/21/2020 | 72,403,796 |
| v1.3 | 4/11/2020 | 1/21/2020 – 4/03/2020 | 87,209,465 |
| v1.4 | 4/13/2020 | 1/21/2020 – 4/10/2020 | 94,671,486 |
| v1.5 | 4/20/2020 | 1/21/2020 – 4/17/2020 | 101,771,227 |
| v1.6 | 4/26/2020 | 1/21/2020 – 4/24/2020 | 109,013,655 |
| v1.7 | 5/04/2020 | 1/21/2020 – 5/01/2020 | 115,929,358 |
| v1.8 | 5/11/2020 | 1/21/2020 – 5/08/2020 | 123,113,914 |

**Table 4**. Breakdown of the most popular languages and the number of associated tweets (v1.8 – May 8, 2020).

| Language | ISO | No. tweets | % total |
|---|---|---|---|
| English | en | 80,698,556 | 65.55% |
| Spanish | es | 13,848,449 | 11.25% |
| Indonesian | in | 4,196,591 | 3.41% |
| French | fr | 3,762,601 | 3.06% |
| Portuguese | pt | 3,451,196 | 2.80% |
| Japanese | ja | 2,897,046 | 2.35% |
| Thai | th | 2,754,627 | 2.24% |
| (undefined) | und | 2,711,649 | 2.20% |
| Italian | it | 1,615,916 | 1.31% |
| Turkish | tr | 1,308,989 | 1.06% |

**Table 5**. Known gaps in the dataset in UTC (v1.8 – May 8, 2020).

| Date | Time |
| --- | --- |
| | |
| 2/1/2020 | 4:00 - 9:00 UTC |
| 2/8/2020 | 6:00 - 7:00 UTC |
| 2/22/2020 | 21:00 - 24:00 UTC |
| 2/23/2020 | 0:00 - 24:00 UTC |
| 2/24/2020 | 0:00 - 4:00 UTC |
| 2/25/2020 | 0:00 - 3:00 UTC |
| 3/2/2020 | Intermittent internet connectivity issues |

## Discussion

We show initial analysis on our collected dataset that verifies that Twitter discourse statistics reflect major events at the time, and leverage Business Insider [29], NBC [30] and CNN [31] released timelines to identify these events of interest during the development of the coronavirus pandemic. In some of these analyses, there is a dip on 3/2/2020 – this was due to internet connectivity failures throughout that specific day. Our discussion is based on analysis done on tweets from release v1.2 (January 21,2020 – March 31, 2020), while the most recent release is v1.8.

## Limitations

We first address several limitations of our dataset. We collect our dataset leveraging Twitter's free stream API which only returns 1% of the total Twitter volume, and the volume of Tweets we collected continues to be dependent on our filter endpoint and network connection [32].

While our dataset is a multi-lingual dataset, containing Tweets from over 67 languages, the keywords and accounts we have and continue to be tracking have been mostly English keywords and accounts. Thus, there is a significant bias in favor of English Tweets in our dataset over Tweets of other languages.

Despite these limitations, our data collection gathers over 1 million Tweets a day from the 1% of Tweets available to us through Twitter's API, and our dataset contains on average 35% non-English Tweets. Our collection begins in late January, capturing Tweets during many major developments – and we plan on continuing collecting Tweets for the foreseeable future.

## Hashtags

We track the frequency of coronavirus-related hashtags, specifically those that contain the substrings "wuhan", "coronavirus" and "covid" throughout our collection period in Figure 1. We can see that while hashtags with the substring coronavirus consistently remains a significantly more heavily used hashtag in our dataset, the hashtag usage spikes when the WHO declared COVID-19 a global public health emergency and on the day that the United States announced the first coronavirus related death [2]. We also do not see hashtags referencing "covid" being used until 2/11/2020, when WHO announced "COVID-19" as the official name for the novel coronavirus. The keyword "wuhan" in hashtags sees steady usage until late February and steadily declines, which reflects the decline in cases in China and the global spread of the virus.

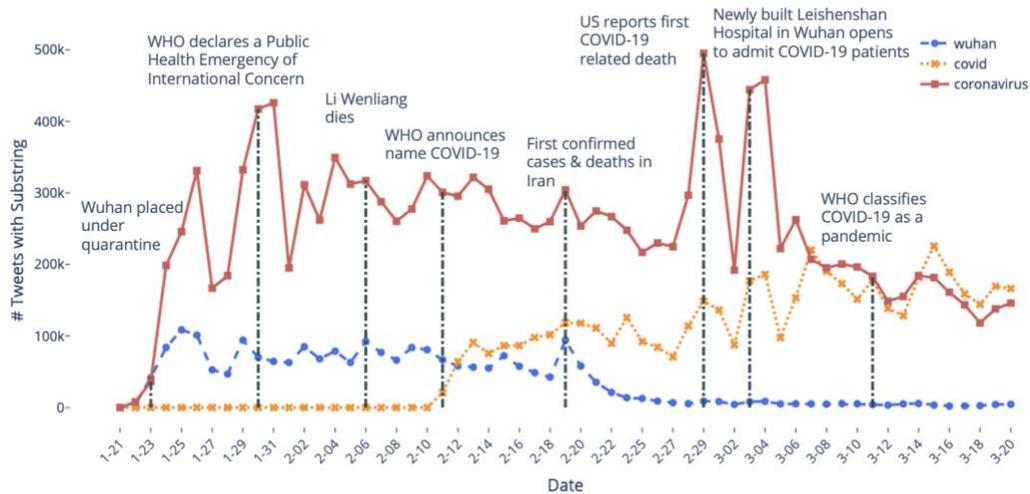

*Figure 1*: Hashtags containing the substrings "wuhan", "covid" and "coronavirus" usage over time.

## Languages

We then examine the percentage of total tweets posted in different languages. Although English is the most prominent language in our dataset, we exclude English from this analysis as to better visualize the activity in Tweets from countries that experienced coronavirus outbreaks earlier in the timeline. In particular, we see that Japanese tweet activity increases steadily after the Diamond Princess is quarantined off the coast of Yokohama, Japan, with a peak around the time when passengers began to disembark the Diamond Princess [33].

There is also a significant spike in Italian tweets when the first case related to COVID-19 is reported in Lodi, Italy, and first death in Veneto [34]. We can also observe a peak in the percentage of Spanish tweets after the first COVID-19 case in Spain is announced on 2/01/2020 [35] and a steady increase in percentage of Spanish tweets after reports of the first COVID-19 related death began to emerge (the death itself occurred on 2/13 but the cause was diagnosed post-mortem) [36].

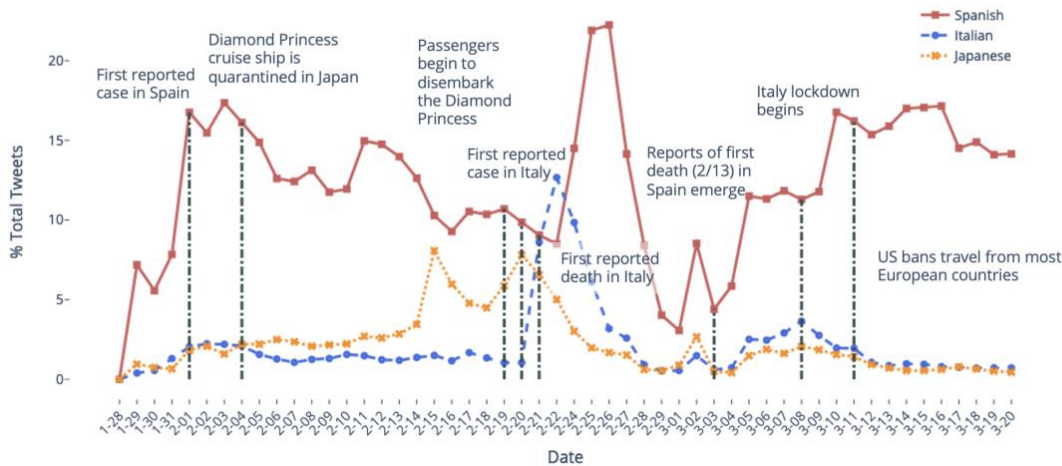

*Figure 2: Tweets in Spanish, Italian and Japanese over time. Our multi-lingual database begins after 1/28/2020.*

### Verified Users

Verified users on Twitter have been identified by Twitter as accounts of public interest and are verified to be authentic accounts [37]. We observe that the verified accounts, which include news sources and political figures, are the most active when major events occur, as can be seen in Figure 3. This is as expected, as influential figures and news sources often weigh in and report on breaking news in real-time, using Twitter as a platform to amplify their messaging. As the United States also drives much of the discourse on Twitter, it is therefore unsurprising that there is a major spike in activity from verified users when the US began to see the first COVID-19-related deaths occur.

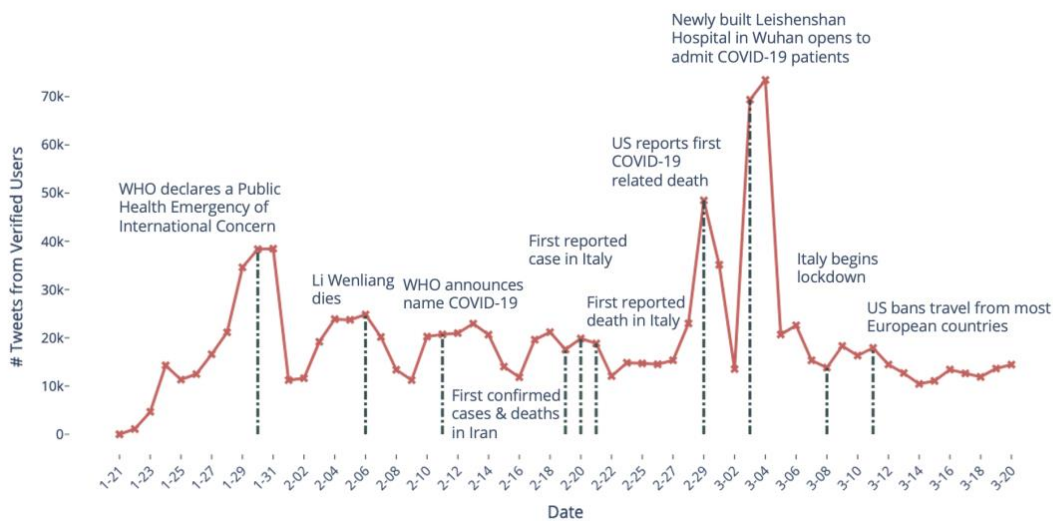

*Figure 3: Number of tweets from verified users over time.*

### Acknowledgements
Data curation: Emily Chen

All authors contributed to the writing of this manuscript.

The authors gratefully acknowledge support from DARPA (contract no. W911NF-17-C-0094).

### Conflicts of Interest
None declared.

### Abbreviations
API: Application Programming Interface
COVID-19: novel coronavirus disease
WHO: World Health Organization
UTC: Coordinated Universal Time